\documentclass[12pt]{article}

\usepackage{graphicx}
\usepackage{dcolumn}
\usepackage{bm}
\newcommand{\beq}{\begin{equation}}
\newcommand{\eeq}{\end{equation}}
\newcommand{\bea}{\begin{eqnarray}}
\newcommand{\eea}{\end{eqnarray}}

\begin{document}
\setlength{\baselineskip}{18pt}
\begin{titlepage}

\begin{flushright}
STUPP-09-204 \\
ICRR-Report-547
\end{flushright}

\vspace*{2mm}
\begin{center}
{\Large\bf 
Higgs Production via Gluon Fusion 
in a Six Dimensional \\
\vspace*{2mm}
Universal Extra 
Dimension Model on $S^2/Z_2$
} 
\end{center}
\vspace{25mm}

\begin{center}
{\large\bf
Nobuhito Maru, Takaaki Nomura$^*$, Joe Sato$^*$ and Masato Yamanaka$^\dag$
%
%
%
}
\end{center}
\vspace{1cm}
\centerline{{\it Department of Physics, Chuo University, Tokyo 112-8551, Japan}}
\vspace*{2mm}
\centerline{{\it $^*$Department of Physics, Saitama University, 
        Saitama, 338-8570, Japan
}}
\vspace*{2mm}
\centerline{{\it
$^\dag$ICRR, University of Tokyo, Kashiwa, Chiba 277-8582, Japan
}}

%
%
\vspace{2cm}
\centerline{\large\bf Abstract}
\vspace{0.5cm}
We investigate Higgs boson production process via gluon fusion at LHC 
in our six dimensional universal extra dimension model compactified on a spherical orbifold $S^2/Z_2$. 
We find a striking result that Higgs production cross section in our model is predicted 
to be 30(10)\% enhancement comparing to the predictions of the Standard Model (the minimal universal extra dimension model) 
for the compactification scale of order 1 TeV. 

\begin{flushleft}
\end{flushleft}
\end{titlepage} 
\newpage

\section{\label{Intro}Introduction}
The universal extra dimension (UED) model is the model 
where all the standard model (SM) fields propagate 
in (TeV)$^{-1}$ extra spatial dimensions \cite{ACD}(See also \cite{Antoniadis}). 
In particular, six dimensional UED models with remarkable features, 
such as electroweak symmetry breaking \cite{ACDH, Hashimoto},  
the prediction of number of generations from anomaly cancellation \cite{DP}, 
and proton stability \cite{ADPY}, 
motivate us to consider such a class of models. 

In our previous paper, 
we have proposed a new six dimensional UED model compactified on $S^2/Z_2$ \cite{MNSY}. 
The gauge group is $SU(3) \times SU(2) \times U(1)_Y \times U(1)_X$ 
and all the SM fields are propagating in the six dimensional bulk. 
A nontrivial background gauge field of $U(1)_X$ is necessary for obtaining massless fermions. 
The Kaluza-Klein (KK) spectrum of all fields were analyzed 
and the lightest KK particles are found to be 1st KK photon and 1st KK gluon at tree level. 
The stability of the lightest KK particles which can be candidates of dark matter 
are ensured by KK parity conservation. 

Large Hadron Collider (LHC) at CERN is about to operate again. 
It is therefore interesting to study collider signatures specific to our model. 
In this Letter, 
we examine Higgs boson production process via gluon fusion in our model. 
Higgs production by gluon fusion is very important 
because it is the dominant production mode at LHC and it has been studied 
in various models beyond the SM \cite{SUSY, Petriello, LH, GH, RS} 
as well as the SM \cite{SM}. 

Before calculating contributions of KK fermions 
 to one-loop effective couplings between Higgs boson and gluons, 
 it is instructive to recall the SM result. 
We parameterize the effective coupling between 
 Higgs boson and gluons as 
\bea 
{\cal L}_{\rm eff} = C_g^{{\rm SM}} \;  h \; G^{a\mu\nu}G^a_{\mu\nu}, 
\label{gfop}
\eea
 where $h$ is a SM higgs boson and 
$G_{\mu\nu}^a$ is a gluon field strength tensor. 
This operator is a dimension six (five) one before (after) electroweak symmetry breaking. 
The coupling is generated by one-loop corrections 
 (triangle diagram) where quarks are running. 
The top quark loop diagram gives the dominant contribution 
 and the coupling $C_g^{{\rm SM}}$ is described in 
 the following instructive form \cite{SM}: 
\bea 
 C_g^{{\rm SM}} = - \frac{m_t}{v} \times 
          \frac{\alpha_s F_{1/2}(4m_t^2/m_h^2)}{8\pi m_t} 
           \times \frac{1}{2},  
\label{gfSM}
\eea 
where, in the right hand side, 
 the first term $- \frac{m_t}{v}$ is top Yukawa coupling, 
 the second term is from the loop integral 
 with the QCD coupling $\alpha_s$ at QCD vertices, 
 the loop function $F_{1/2}(\tau)$ given by (for $\tau \geq 1$) 
\bea
 F_{1/2}(\tau) &=& -2 \tau \left( 
   1+ \left( 1 - \tau \right) 
   \arcsin^2(1/\sqrt{\tau}) \right) \nonumber \\
   &\to& -\frac{4}{3}~~{\rm for}~\tau \gg 1 ,  
\label{loopfunc}
\eea
 and $1/2$ is a QCD group factor (Dynkin index). 
Mass of the fermion (top quark) running in the loop 
 appears in the denominator in the second term of (\ref{gfSM}), 
 which is canceled with top quark mass from Yukawa coupling. 
It is well-known that in the top quark decoupling limit, 
 namely top quark mass $m_t$ is much heavier 
 than Higgs boson mass $m_h$, 
 $F_{1/2}$  becomes a constant and 
 the resultant effective coupling becomes independent 
 of $m_t$ and $m_h$.

\section{Calculation of Gluon Fusion Amplitude}

A calculation of KK mode contributions to gluon fusion in our model  
is completely analogous to the top loop correction in the SM case. 
To carry out the calculation, we need to know top Yukawa coupling constant 
and KK mass spectrum of top quark in our model in a mass eigenstate. 
The relevant top Yukawa coupling and mass terms after electroweak symmetry breaking 
are given by
\begin{eqnarray}
{\cal L} 
&\supset& \sum_{lm} \left[
M_l \bar{T}^{lm} T^{lm} 
-M_l \bar{Q}^{lm} Q^{lm} 
+ \frac{m_t}{v} \left( 
 \bar{T}^{lm} H^{00} Q^{lm} 
+ \bar{Q}^{lm} H^{00} T^{lm} \right) 
+ m_t \left( 
 \bar{T}^{lm} Q^{lm} 
+ \bar{Q}^{lm} T^{lm} \right) 
\right] 
\label{massYukawa}
\nonumber \\
\end{eqnarray}
where 
$m_t$ is top quark mass, 
and $v$ is a vacuum expectation value of Higgs field $H^{00}(x)$. 

The mass terms for KK modes can be written down in a matrix form by using Dirac fermion
\begin{eqnarray}
\left(
\begin{array}{cc} 
\bar{T}^{lm} & \bar{Q}^{lm} 
\end{array}
\right)
\left( 
\begin{array}{cc} 
M_l & m_t \\ 
m_t & -M_l 
\end{array}
\right)
\left( 
\begin{array}{c} 
T^{lm} \\ 
Q^{lm} 
\end{array}
\right), 
\end{eqnarray}
which is diagonalized by the change of basis
\begin{eqnarray}
\left(
\begin{array}{c} 
T^{lm} \\ 
Q^{lm} 
\end{array}
\right) =
\left( 
\begin{array}{cc} 
\gamma_5 \cos \alpha_l & \sin \alpha_l \\ 
-\gamma_5 \sin \alpha_l  & \cos \alpha_l 
\end{array}
\right)
\left( 
\begin{array}{c} 
T'^{lm} \\ 
Q'^{lm} 
\end{array}
\right)
\end{eqnarray}
where $\tan 2\alpha_l = m_t/M_l$. 
Rewriting (\ref{massYukawa}) in terms of mass eigenstates $T'^{lm}$ and $Q'^{lm}$, we find that  
 top KK mass eigenvalue is $m_{t}^{(l)} = \sqrt{\frac{l(l+1)}{R^2} + m_t^2}$ 
 and top Yukawa coupling is $- (m_t \sin 2\alpha_l) /v = - m_t^2/(v m_t^{(l)})$, 
respectively \cite{MNSY}. 
Making use of this information, 
the KK mode contributions in our model are found to be  
\bea
{\cal L}_{{\rm eff}} &=& C_g^{{\rm KK(UED}_2)} \; 
  h \; G^{a\mu\nu} G^a_{\mu\nu}
\eea
where
\bea 
C_g^{{\rm KK(UED}_2)} &=& - \sum_{l=1}^\infty n(l)
 \left[
 \frac{m_t}{v} \frac{m_t}{m_t^{(l)}} \times 
 \frac{\alpha_s F_{1/2}(4 (m_{t}^{(l)})^2/m_h^2)}
 {8\pi m_{t}^{(l)}} \frac{1}{2} \right] \times 2 \nonumber \\
&=& \frac{\alpha_s}{\pi v} \frac{m_t^2}{m_h^2} \sum_{l=1}^\infty 
\left[ 
(2l+1) 
\left\{
1 + \left(1 - \frac{4(m_t^{(2l)})^2}{m_h^2} \right) \arcsin^{2} \left(\frac{m_h}{2m_t^{(2l)}} \right)
\right\} 
\right. \nonumber \\
&& 
+ 
\left. 
\frac{\alpha_s}{\pi v} \frac{m_t^2}{m_h^2} \sum_{l=1}^\infty 
(2l-1) 
\left\{
1 + \left(1 - \frac{4(m_t^{(2l-1)})^2}{m_h^2} \right) \arcsin^{2} \left(\frac{m_h}{2m_t^{(2l-1)}} \right)
\right\}
\right] 
\nonumber \\ 
&\simeq& 
 \frac{\alpha_s}{6 \pi v} \sum_{l=1}^{\infty}
 \left[ 
\frac{(2l+1)m_t^2}{\frac{2l(2l+1)}{R^2}+m_t^2} 
+ 
\frac{(2l-1) m_t^2}{\frac{2l(2l-1)}{R^2}+m_t^2} 
\right]
\eea
where ``UED$_{2(1)}$" denotes our 6D UED model on $S^2/Z_2$ \cite{MNSY} 
(5D UED model on $S^1/Z_2$ \cite{ACD}, which will be discussed later), respectively. 
A factor ``2" is multiplied in the second line,  
since the degrees of freedom of 6D fermion are doubled compared with the SM case. 
In the second and the third equalities, 
the mode sum is decomposed into even or odd number term of $l$ 
since the degeneracy $n(l)$ with respect to $m$ is different, 
{\it e.g.} $n(l)= l+1(l)$ for $l:$ even (odd) \cite{MNSY}. 
The limit $m_h^2, m_t^2 \ll (1/R)^2$ have been taken 
in the last line to simplify the results. 
As expected from the dimensional analysis, 
the mode sum is logarithmically divergent.  
Also, note that the KK mode contribution is constructive 
 against the top quark contribution in the SM. 

It is interesting to compare our result with 
 that in the minimal UED model on $S^1/Z_2$ \cite{Petriello}, 
 where the KK mode mass spectrum and Yukawa couplings 
 are given by 
 $M_n = \sqrt{(n/R)^2 + m_t^2}$  
 and $-(m_t^2/M_n v)$, respectively. 
In this case, we find the effective coupling as
\bea 
{\cal L}_{{\rm eff}} &=& C_g^{{\rm KK(UED}_1)} \; 
  h \; G^{a\mu\nu} G^a_{\mu\nu} 
\eea
where
\bea
C_g^{{\rm KK(UED}_1)} 
= 
 -\sum_{n=1}^\infty 
 \left[
 \frac{m_t}{v} \frac{m_t}{M_n}
 \times 
 \frac{\alpha_s F_{1/2}(4 M_n^2/m_h^2)}{8 \pi M_n} 
 \times \frac{1}{2} \right]  \times 2 
\simeq 
 \frac{\alpha_s}{6 \pi v}
  \sum_{n=1}^\infty \frac{m_t^2}{(n/R)^2}  
\label{gfS1Z2UED}
\eea
where we have taken the limit $m_h^2,~m_t^2 \ll (1/R)^2$ again to simplify the result. 
This KK mode contribution is finite and constructive to 
 the top quark one in the SM. 

As we have shown, the KK mode loop contribution to 
 the effective coupling between Higgs boson and 
 gluons is constructive similar to 
 the top quark loop contribution in the SM. 
This fact leads to remarkable effects on Higgs boson 
 search at the LHC.  
Since the main production process of Higgs boson at the LHC 
 is through gluon fusion, so that the deviation 
 of the effective coupling between Higgs boson and gluons from the SM and other model's predictions 
 directly affects the Higgs boson production cross section. 

Let us consider the ratio of the Higgs boson production cross section 
 in the UED model on $S^2/Z_2$ and $S^1/Z_2$ to the SM one, 
 which is described as 
\bea 
\Delta \equiv  \frac{\sigma(gg \to h;~{\rm SM+KK})}
       {\sigma(gg \to h;~{\rm SM})}  =
  \left( 1 +  \frac{C_g^{{\rm KK(UED}_{2(1)})}}{C_g^{{\rm SM}}} \right)^2. 
\eea
\begin{figure}
\begin{center}
\includegraphics[scale=0.9]{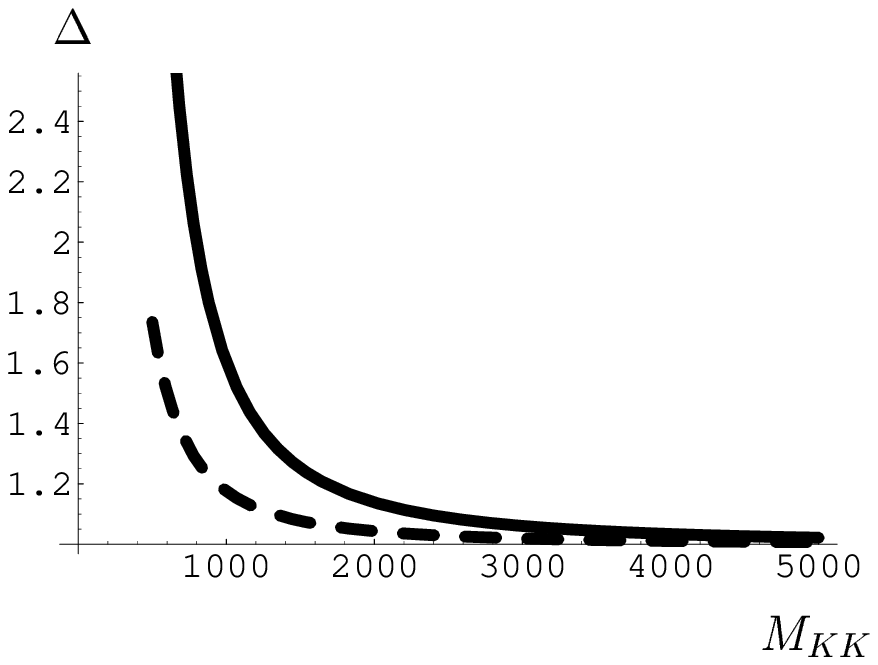}
\quad
\includegraphics[scale=0.9]{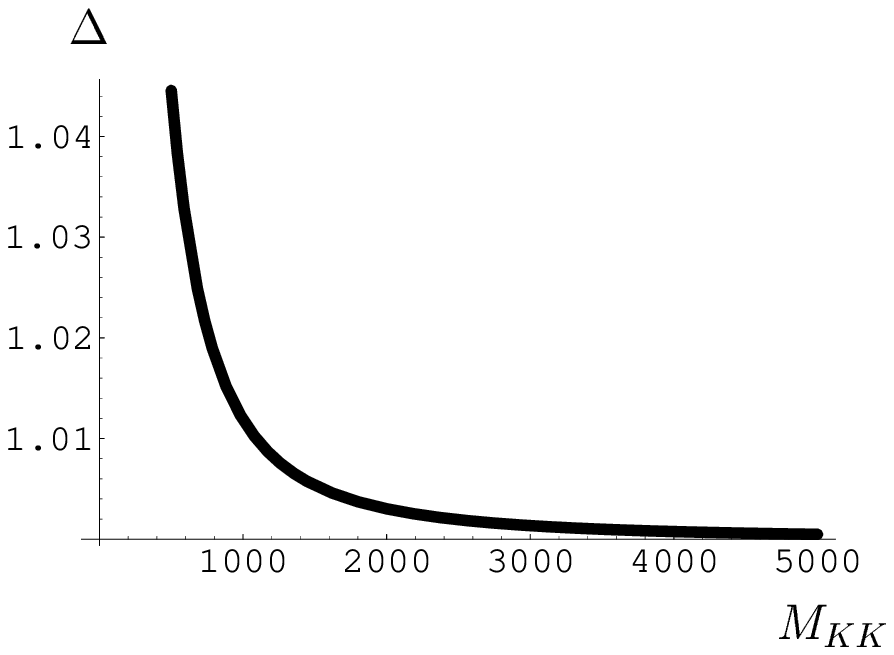}
\caption{\label{plot} 
The left plot represents the ratio of the Higgs boson production cross section via gluon fusion 
 in 6D UED on $S^2/Z_2$(bold line), 5D UED on $S^1/Z_2$(dashed line) and that in the SM 
 as a function of the first KK mass in a unit of GeV ($\sqrt{2}/R$ for $S^2/Z_2$ and $1/R$ for $S^1/Z_2$). 
The right plot represents the ratio of the Higgs boson production cross section via gluon fusion 
 in 6D UED on $S^2/Z_2$ and $T^2/Z_2$ 
 as a function of the first KK mass $M_{KK}$ in a unit of GeV ($\sqrt{2}/R$ for $S^2/Z_2$ and $1/R$ for $T^2/Z_2$). 
The vertical axis denotes $\Delta$ defined in the text and 
$\Delta=1.0$ corresponds to the SM (6D UED on $T^2/Z_2$) prediction in the left (right) plot. 
Higgs mass is taken to be 120 GeV. 
}
\end{center}
\end{figure}
The numerical result of this ratio as a function of the first KK mass scale 
is shown in the left hand side of Fig.~\ref{plot} where the plot is calculated 
by using exact expressions of $C_g^{{\rm SM}}, C_g^{{\rm KK(UED}_{2(1)})}$ not approximated ones. 
The bold (dashed) line corresponds to the UED model on $S^2/Z_2(S^1/Z_2)$ with fixed Higgs mass to be 120 GeV. 
The horizontal line $\Delta=1$ is the SM prediction. 
We have also calculated the cases where Higgs mass is $m_h=150, 180$ GeV, 
but the results do not change so significantly 
since the decoupling limit $4 M_l^2/m_h^2 \gg 1$ are well satisfied. 
The KK fermion contribution is constructive 
 and the Higgs production cross section is increased in the UED scenario 
in contrast to the case of little Higgs \cite{LH} or gauge-Higgs unification \cite{GH}. 
The present UED model on $S^2/Z_2$ gives rise to 
more enhanced Higgs production cross section 
than that of the minimal UED model on $S^1/Z_2$. 
This is very natural because the number of KK particles is larger. 
For the compactification scale around 1 TeV, the KK fermion contribution of our model 
 is sizable and the production cross section is more enhanced by around 30(10)\% 
than the SM (minimal UED on $S^1/Z_2$) prediction. 
Thus, we have found that our model predicts a remarkable collider signature of Higgs production at LHC. 
We expect that our prediction will be soon verified by the forthcoming experiment. 

We further compare our results with that of 6D UED model compactified on $T^2/Z_2$. 
The effective coupling of such an example can be easily obtained by replacing the mass spectrum 
of KK top quarks $m_t^{(l)}$ in $C_g^{{\rm KK(UED}_2)}$ with $m_t^{(l,m)} = \sqrt{\frac{l^2+m^2}{R^2} + m_t^2}$ 
in the simplest case with identical radii. 
The result is shown in the right hand side of Fig. \ref{plot}.
We can see that the deviation is of order a few percent. 
This result can be understood as follows. 
The gluon fusion amplitude is logarithmically divergent, in other words, depends on the UV physics logarithmically. 
However, the UV physics is not affected so much by the way to compactify since the compactification is an IR physics. 
Therefore, the predictions of 6D UED models on $S^2/Z_2$ and $T^2/Z_2$ are almost the same.

In our analysis, we have summed only the first five KK mode contributions.  
The reason is the following. 
Our model is a six dimensional model, 
so the gluon fusion amplitude is logarithmically divergent as mentioned earlier. 
More specifically, the mode sum behaves as $\log(\Lambda R)$ with the cutoff scale $\Lambda$. 
Therefore, one might worry about the cutoff dependence of the result. 
We can find an upper bound for the cutoff scale by using naive dimensional analysis, 
as discussed in \cite{ACD}. 
A loop expansion parameter $\varepsilon$ of six dimensional theories is given by 
\begin{eqnarray}
\varepsilon = \frac{\pi^3}{2(2\pi)^6} g_6^2 \Lambda^2 = \frac{\alpha}{8\pi}(R \Lambda)^2 
\end{eqnarray}
where $g_6$ is a gauge coupling constant in six dimensional gauge theories. 
The cutoff scale is introduced to make $\varepsilon$ dimensionless. 
$\alpha \equiv g_4^2/(4\pi)(g_4: {\rm 4D~gauge~coupling~constant})$. 
Requiring that our theory is perturbative at the cutoff scale, $\varepsilon \le 1$, 
we obtain
\begin{eqnarray}
R \Lambda \le \sqrt{\frac{8\pi}{\alpha}}. 
\end{eqnarray}
The most stringent bound is found for the case that 
our 4D effective theory becomes strong coupling at the cutoff scale $\alpha \simeq 1$. 
Thus, we finally obtain the upper bound of the cutoff scale
\begin{eqnarray}
\Lambda \le \sqrt{8\pi}/R \simeq 5/R. 
\end{eqnarray}
Above this upper bound, our theory is not reliable since the perturbativity is lost 
and should be replaced with some underlying theory. 
Thus, taking into account the first five KK modes is appropriate in the viewpoint of our effective theory. 

Here is a comment on collider physics. 
If we take into account the Higgs boson decay process, 
our result holds true for the case where Higgs mass is heavier than around 150 GeV. 
In that situation, Higgs boson will mainly decay into $W^+ W^-$ pair via the SM vertex at tree level. 
Therefore, $\Delta$ is unchanged. 
However, if Higgs boson mass is lighter than 150 GeV, 
the most promising discovery mode is two photon decay process. 
This process is also given by one-loop triangle diagram and 
KK modes of top quark and W-boson contribute.
The analysis of Higgs boson decay into two photons is beyond the scope of this paper 
and is left for future work \cite{2photon}. 

\section{Summary}

In summary, 
we have investigated the main Higgs production process via gluon fusion at LHC 
in our 6D UED model compactified on $S^2/Z_2$. 
This is the first result in 6D UED models. 
Higgs production cross section in our model has 30 (10)\% enhancement 
comparing with the prediction of the SM (minimal UED model on $S^1/Z_2$) 
for the compactification scale of order 1 TeV. 
We have also also compared our results with predictions of the 6D UED model 
compactified on $T^2/Z_2$ with identical radii. 
The result is found to be almost the same, namely almost independent of the way of compactification. 

Our results are affected by UV physics because of the logarithmic cutoff scale dependence. 
In comparing our results with experimental ones, we have to care about this. 
Although we have such an subtle issue, we expect that our remarkable prediction will be soon verified 
by a forthcoming LHC experiment. 

\vspace*{2mm}

\begin{center}
{\bf Acknowledgments}
\end{center}
The authors were supported in part by the Grant-in-Aid for Scientific Research 
of the Ministry of Education, Science and Culture, 
[No.18204024(N.M.), No.19010485(T.N.), No.20025001, 20039001, 20540251(J.S.), No.20007555(M.Y.)].

\appendix
\section{Brief review of our model}

In this appendix, we briefly review our model \cite{MNSY}.  
Our model is defined on the six-dimensional spacetime $M^6$ 
which is compactified on a two-sphere orbifold 
$M^6=M^4 \times S^2/Z_2$.
We denote the coordinate of $M^6$ by $X^M=(x^{\mu},y^{\theta}=\theta,y^{\phi}=\phi)$, 
where $x^{\mu}$ and $\{\theta,\phi \}$ are the $M^4$ coordinates and are the $S^2/Z_2$ spherical coordinates, respectively.
On the orbifold, the point $(\theta,\phi)$ is identified with $(\pi-\theta,-\phi)$.
The spacetime index $M$ runs over $\mu \in \{0,1,2,3 \}$ and $\alpha \in \{\theta,\phi\}$.
The metric of $M^6$ can be written as 
\begin{eqnarray}
g_{MN} = \left(
\begin{array}{cc} 
\eta_{\mu \nu} & 0 \\ 
0 & -g_{\alpha \beta} \\
\end{array}
\right), 
\end{eqnarray}
where $\eta_{\mu \nu}= diag(1,-1,-1,-1)$ and $g_{\alpha \beta}= diag(R^{2}, R^{2}\sin^{2} \theta)$ 
are metric of $M^4$ and $S^2/Z_2$ respectively, and $R$ denotes the radius of $S^2$.
We define the vielbein $e^{M}_{A}$ 
that connects the metric of $M^6$ and that of 
the tangent space of $M^6$, denoted by $h_{AB}$, as $g_{\scriptscriptstyle MN}=e_M^{A} e_N^B h_{AB}$. 
Here $A=(\mu,a)$, where $a$ $\in$ $\{ 4,5 \}$, is the index for the coordinates of tangent space of $M^6$. 
%
%

%
We introduce, in this theory, a gauge field $A_{M}(x,y)=(A_{\mu}(x,y),A_{\alpha}(x,y))$, 
SO(1,5) chiral fermions $\Psi_{\pm}(x,y)$,
 and complex scalar fields $\Phi(x,y)$.
The SO(1,5) chiral fermion $\Psi_{\pm}(x,y)$ is defined by the action of SO(1,5) chiral operator $\Gamma_7$,
 which is defined as
\begin{equation}
\Gamma_7 = \gamma_5 \otimes \sigma_3,
\end{equation}
where $\gamma_5$ is SO(1,3) chiral operator and $\sigma_i(i=1,2,3)$ are Pauli matrices.
The chiral fermion $\Psi_{\pm}(x,y)$ satisfies 
\begin{equation}
\Gamma_7 \Psi_{\pm}(x,y)= \pm \Psi_{\pm}(x,y)
\end{equation}
and is obtained by acting the chiral projection operator of SO(1,5), $\Gamma_{\pm}$, on 
Dirac fermion $\Psi(x,y)$, where 
$\Gamma_{\pm}$ is defined as 
\begin{equation}
\Gamma_{\pm} = \frac{1 \pm \Gamma_7}{2}.
\end{equation}
We can also write $\Psi_{\pm}(x,y)$ in terms of SO(1,3) chiral fermion $\psi$ as 
\begin{eqnarray}
\label{chiralR}
\Psi_+ = 
\left(
\begin{array}{c} 
\psi_R \\ 
\psi_L 
\end{array}
\right), \quad
\label{chiralL} 
\Psi_- =
\left( 
\begin{array}{c} 
\psi_L \\ 
\psi_R 
\end{array}
\right), 
\end{eqnarray}
where $\psi_{R(L)}$ is a right(left)-handed SO(1,3) chiral fermion. 
%
%
The boundary conditions for each 
field can be defined as 
\begin{eqnarray}
\label{BC0}
\Phi(x,\pi-\theta,-\phi) &=& \pm \Phi(x,\theta,\phi), \\
\label{BC1}
A_{\mu}(x,\pi-\theta,-\phi) &=& A_{\mu}(x,\theta,\phi), \\
\label{BC2}
A_{\theta,\phi}(x,\pi-\theta,-\phi) &=& -A_{\theta,\phi}(x,\theta,\phi), \\
\label{BC3}
\Psi(x,\pi-\theta,-\phi) &=& \pm \gamma_5 \Psi(x,\theta,\phi)
\end{eqnarray}
by requiring the invariance of a six-dimensional action under the $Z_2$ transformation.

The action of the gauge theory is written as 
\bea
\label{6Daction}
S &=& \int dx^4 R^2 \sin \theta d \theta d \phi 
\left(
\bar{\Psi}_{\pm} i \Gamma^{\mu} D_{\mu} \Psi_{\pm} + \bar{\Psi}_{\pm} i \Gamma^{a} e^{\alpha}_{a} D_{\alpha} \Psi_{\pm}  
- \frac{1}{4 g^2} g^{MN} g^{KL} Tr[F_{MK} F_{NL}] 
\right. \nonumber \\
&& \left. 
+(D^M \Phi)^* D_M \Phi -V(\Phi) 
-y \bar{\Psi}_{\pm} \Phi \Psi_{\mp} \right) , 
\eea
where $F_{MN}= \partial_M A_N -\partial_N A_M -[A_M,A_N]$ is the field strength, 
$D_M$ is the covariant derivative including a spin connection, $V(\Phi)$ is the scalar potential term, 
$y$ is Yukawa coupling constant, 
and $\Gamma_A$ represents the 6-dimensional Gamma matrices. 
%
%

%
As discussed in \cite{MNSY}, 
the positive curvature of an extra-space $S^2$ gives mass to fermions in four-dimensional spacetime, 
%
%
%
we then introduce a background gauge field 
$A^B_{\phi} = \hat{Q} \cos \theta$ \cite{S2}
where $\hat{Q}$ is a charge of some U(1) gauge symmetry, in order to cancel the mass from the curvature 
and to obtain massless fermions in four-dimensional spacetime.
Indeed, $A^B_{\phi}$ cancel the spin connection term for the upper(lower) component SO(1,3) fermion in Eq.~(\ref{chiralR})   
if the fermion has the charge $\hat{Q}=+(-) \frac{1}{2}$ and the upper(lower) component gets a massless Kaluza-Klein mode.



Here, we focus on the top quark sector in our model, which is relevant for calculation of the gluon fusion process. 
Consider the standard model gauge group with an extra U(1)$_X$ gauge symmetry
, i.e. $G = SU(3) \times SU(2) \times U(1)_Y \times U(1)_X$. 
%
As mentioned earlier, 
we must introduce the extra U(1) to obtain all the massless chiral SM fermions in $M^4$. 
%
%
We then assign the extra U(1) charge $\hat{Q}=\frac{1}{2}$ to these fermions as the simplest case 
 in which all massless SM fermions appear in four-dimensional spacetime \cite{MNSY}. 
%

The action of top quark sector in our model is written as 
\begin{eqnarray}
\label{action6D}
S_{6D} &=& \int dx^4 R^2 \sin \theta d \theta d \phi 
\left[
\bar{Q}_- i \Gamma^M D_M Q_- 
+ \bar{T}_+ i \Gamma^M D_M T_+ 
+ (  y_t \bar{Q}_- T_+ H + \textrm{h.c}) 
 \right], 
\end{eqnarray}
where $Q(x,y)_-, T(x,y)_+$ belong to representations of $SU(3) \times SU(2) \times U(1)_Y$ 
which are the same as the left-handed quark doublet, right-handed top quark. 
The Higgs field $H(x,y)$ does not have a $U(1)_X$ charge and is even under the $Z_2$ action 
so that Yukawa coupling terms are allowed.

The boundary conditions of these fermions for 6D chirality and $Z_2$ parity 
should be imposed so as to realized massless fermions as, 
\bea
Q_-(x, \pi - \theta, -\phi) &=& -\gamma_5 Q_-(x, \theta, \phi), \\
T_+(x, \pi - \theta, -\phi) &=& +\gamma_5 T_+(x, \theta, \phi). 
\eea
The 4D action is obtained by integrating the Lagrangian over $S^2/Z_2$ coordinate.
\bea
S_{4D} &=& \int d^4x 
\left[
\sum_{lm} \left(
\bar{Q}^{lm}(i \partial \!\!\!/ - M_l) Q^{lm} + \bar{T}^{lm}(i \partial \!\!\!/ + M_l) T^{lm} 
\right) \right. \nonumber \\
&& \left. 
+ \sum_l \frac{[1+(-1)^l]^2}{4} 
\left(
\bar{Q}^{l0}(i \partial \!\!\!/ - M_l) Q^{l0} + \bar{T}^{l0}(i \partial \!\!\!/ + M_l) T^{l0} 
\right) \right. \\
&& \left. 
+ \bar{Q}_L^{00}i \partial \!\!\!/  Q_L^{00} + \bar{T}_R^{00}i \partial \!\!\!/ T_R^{00} 
+ ( y_t \bar{T}^{lm} H^{00} Q^{lm}  + {\rm h.c.} )
\right] \nonumber
\eea
where $M_l = \sqrt{l(l+1)}/R$.



\end{document}